\documentclass[aps,pre,twocolumn,showpacs,floatfix]{revtex4-1}
\usepackage{graphicx}
\usepackage{graphics}
\usepackage{amssymb}
\usepackage{hyperref}
\usepackage{latexsym}
\usepackage{amsmath}
\usepackage{textcomp}
\usepackage{amsfonts}
\usepackage{color}
\usepackage{bm}

\newcommand{\Dmat}{\tilde{\mathcal D}}
\newcommand{\vBZ}{v_{\rm BZ}}

\newcommand{\bR}{{\bf R}}

\begin{document}

 \title{Statistics of non-affine defect precursors: tailoring defect densities in colloidal crystals using external fields}

\author{Saswati Ganguly$^1$, Surajit Sengupta$^{2}$, Peter Sollich$^{3}$ 
}
\affiliation{$^1$ Indian Association for the Cultivation of Science, 2A\&2B Raja S. C. Mullick Road, Jadavpur, Kolkata 700032, India \\
$^2$ TIFR Centre for Interdisciplinary Sciences, 21, Brundavan Colony, Narsingi, Hyderabad 500075, India \\
$^3$ King's College London, Department of Mathematics, Strand, London WC2R 2LS, U.K.\\ 
}

\begin{abstract}
Coarse-graining atomic displacements in a solid produces both local affine strains and ``non-affine'' fluctuations. Here we study the equilibrium dynamics of these coarse grained quantities to obtain space-time dependent correlation functions. We show how a subset of these thermally excited, non-affine fluctuations act as precursors for the nucleation of lattice defects and suggest how defect probabilities may be altered by an {\it experimentally realisable} ``external'' field conjugate to the global non-affinity parameter. Our results are amenable to verification in experiments on colloidal crystals using commonly available holographic laser tweezer and video microscopy techniques, and may lead to simple ways of controlling the defect density of a colloidal solid.  
\end{abstract}

\maketitle

\section{Introduction}
\label{intro}
While a large body of work has accumulated over many decades on the physics of crystal defects~\cite{Baluffi}, the microscopic causes of defect nucleation and yielding in solids remain active areas of recent research~\cite{schall-nat,PNAS}. A small external stress on a crystalline solid at non-zero temperatures affects atomic configurations in two ways: (1) an affine deformation characterised by the elastic strain and (2) a modification of the relative probabilities of thermally excited lattice defects~\cite{MTE,CL,SS1}. Within a linear response picture~\cite{CL}  local strain fluctuations measured from particle coordinates determine the elastic moduli of the solid, which in turn govern the magnitude of the affine response~\cite{SS1,zahn,zhang,harm-colloid,kers1}. It is therefore natural to ask the complementary question viz.\ fluctuations of which quantity, derivable {\it solely} from the configuration of the atoms, measure the susceptibility of a crystalline solid to creation of defects? In this paper, we pursue this issue by extending and generalising an approach introduced in~\cite{sassy} based on coarse-graining of atomic displacements. Soft, precursor fluctuations which give rise to defects appear as a natural outcome of this coarse-graining process. We explore some of the interesting consequences of this connection -- such as the ability to engineer equilibrium defect concentrations, at least in a colloidal crystal~\cite{colloids}, by subtly altering the statistical weights of these precursors using dynamic light fields~\cite{HOT}.   
 
Consider a system consisting of $i = 1 \dots N$ particles with instantaneous positions $\{{\bf r}\}$ vibrating about a set of reference coordinates $\{{\bf R}\}$. To begin, we first elevate a measure of non-affinity
 introduced in~\cite{falk} to identify elastic heterogeneities in sheared amorphous solids, to the role of a fully-fledged thermodynamic, collective coordinate. This variable, {\color{black} $X = N^{-1}\sum_{i=1}^{N} \chi({\bf R}_{i})$}, a scalar functional of both the instantaneous and the reference coordinates, measures the magnitude of non-elastic deviations of the positions of all particles away from their reference configuration coarse-grained over a reference volume $\Omega$. {\color{black}The local $\chi({\bf R}_{i})$ is a function of the instantaneous and reference positions ${\bf r}$ and ${\bf R}$ of the particles in the neighborhood $\Omega$ of a given particle $i$ with reference particle position ${\bf R}_{i}$}. We had earlier obtained the equilibrium statistics of {\color{black} of $\chi$ (spatial dependence suppressed for brevity)}, in crystals~\cite{sassy} at finite temperatures.  We had shown that under an external stress ${\bf \Sigma}$, particles undergo both affine and non-affine deviations, with {\color{black}$\chi$} always increasing as ${\bf \Sigma}^2$ within the harmonic approximation. 

Here, we first extend the work reported in~\cite{sassy} to {\it time-dependent} correlation functions for {\color{black}$\chi$} and strains at zero stress. This part of our treatment is similar in spirit to that of~\cite{vine} where the dynamical {\em density} correlations are analysed in terms of a sum over harmonic degrees of freedom.  The relaxation time of an observable arises, within this approximation, from the de-phasing of incoherent harmonic oscillations. While our treatment is perfectly general and is applicable to any solid in any dimension for which $\{{\bf R}\}$ and the interactions 
are known, we present results for the two dimensional triangular lattice. Next, an analysis of the vibrational modes contributing to non-affine distortions of $\Omega$ reveals that most of {\color{black}$\chi$} arises from two degenerate non-affine displacements which tend to replace four $6$ coordinated particles with two pairs of particles with $5$ and $7$ neighbours: an incipient, or precursor, dislocation-anti-dislocation pair. Having obtained time dependent correlation functions of {\color{black}$\chi$}, we identify a field, $h_X$, thermodynamically conjugate to the global non-affinity parameter $X$ and argue that the response of the solid to an application of this field would change the magnitude of $X$ (and {\color{black}$\chi$}) and hence the probability of these localised precursor fluctuations that give rise to dislocation pairs. Since $h_X$ couples linearly, {\color{black}$X$} can both increase or decrease depending on the sign of $h_X$. Further, unlike stress, $h_X$ changes defect probabilities directly {\it without} introducing affine strains and associated spatial anisotropies, simplifying the study of their statics and dynamics. Finally, since {\color{black}$X$} is given entirely in terms of the particle coordinates and the $\{{\bf R}\}$ which appear as constant parameters, $h_X$ can be directly introduced into the Hamiltonian and the dynamics of system calculated using standard molecular dynamics~\cite{UMS}. For a colloidal solid, it is even possible to apply $h_X$ in the laboratory using laser tweezers~\cite{HOT, tweeze}, a point that we elaborate later. We explore the effect of $h_X$ on the equilibrium statistics and dynamics of {\color{black}$X$(and $\chi$)} and of the individual defect precursor modes.  

Before we end this introduction, we mention an aspect of our work that makes it relevant to some recent studies of the mechanical response of soft amorphous solids~\cite{cates-evans} and glasses~\cite{falk-review}. In such solids, it is impossible to define the kinds of defect configurations encountered in crystals such as vacancies, dislocations, stacking faults or grain boundaries~\cite{rob}. However, given any reference configuration and a set of particle coordinates, $\chi$ and much of everything else discussed in our work can still be defined and computed. The precursor modes in this case should be related to non-affine droplet fluctuations, which have been extensively studied in recent years~\cite{argon,falk,quad,lemat, barat, zero-T, itamar, schall, schall1, schall2}. Unlike crystals, however, the identification of defect precursors with actual defects is much more problematic for amorphous solids due to the lack of a simple and unique reference $\{{\bf R}\}$ and is, therefore, a subject of ongoing lively debate~\cite{manning1,manning2,itamar2}. We hope that some of the ideas discussed here may be useful in illuminating this issue. We return to a discussion of this point briefly later.

The rest of the paper is organised as follows. In section~\ref{sec:1} we set up the calculation and define the coarse-graining process used to calculate spatio-temporal correlation functions. Parts of this calculation have previously appeared in~\cite{sassy}, but we include the relevant aspects here for completeness and to make the paper self-contained. In ~\ref{sec:2} we present our results for the time dependent $\chi$ and strain fluctuations in the two-dimensional triangular lattice. In ~\ref{sec:3} we identify defect precursors and obtain their statistics. We also introduce the non-affine field $h_X$ and study its effect on these precursor fluctuations. In section~\ref{sec:4} we suggest how $h_X$ may be produced in the laboratory using laser tweezers~\cite{HOT}. We discuss our results and conclude by giving indications of future directions in section~\ref{sec:5}.

\section{Coarse graining and dynamic correlation functions}
\label{sec:1}
On application of an external stress or as a result of thermal fluctuations, particles $i$ within a solid undergo displacements ${\bf u}_{i} = {\bf r}_{i}-{\bf R}_{i}$ away from some chosen reference configuration ${\bf R}_{i}$ to their displaced positions ${\bf r}_{i}$. In a homogeneous solid at vanishing temperature, such displacements are affine, implying that they can be expressed as ${\bf u}_{i} = \mathsf{D}{\bf R}_{i}$, where $\mathsf{D} = \mathsf{K}^{-1}\bf{\Sigma}$ is the deformation tensor related to the external stress $\bf{\Sigma}$ via the tensor of elastic constants $\mathsf{K}$. To derive the closest approximation to this simple zero temperature scenario in the presence of thermal fluctuations we proceed as follows \cite{sassy}. 

Consider a neighbourhood, $\Omega$, larger than the unit cell, around a central particle labelled $0$ 
consisting of $N_\Omega$ particles $i$ within a cut-off distance $R_\Omega$ in a $d$ dimensional lattice.  The reference, zero temperature lattice configurations are labelled by ${\bf R}_{i = 0\dots N_\Omega}$ while the fluctuating atom positions are ${\bf r}_{i = 0\dots N_\Omega}$. The particle displacements are then {\color{black} as before} ${\bf u}_i = {\bf r}_{i} - {\bf R}_{i}$. Now define relative displacements, ${\bf \Delta}_{i} = {\bf u}_i-{\bf u}_0 = {\bf r}_{i} - {\bf r}_{0} -({\bf R}_{i} - {\bf R}_{0})$ of particle $i$ compared to particle $0$. The ``best fit'' \cite{falk} coarse-grained local deformation tensor ${\mathsf D}$ is the one that minimizes  $\sum_i
[{\bf \Delta}_{i} - {\mathsf D}({\bf R}_{i} - {\bf R}_{0})]^2$ with the non-affinity parameter $\chi$ being the (positive definite) minimum value of this quantity.

In~\cite{sassy} we showed that the result of this minimisation procedure may be expressed as a projection of the particle displacements ${\bf \Delta}_i$ into mutually orthogonal subspaces as defined by two projection operators ${\mathsf P}$ and ${\mathsf R}{\mathsf Q}$. 
In terms of these,
 $\chi= {\bf \Delta}^{\rm T}{\mathsf P}{\bf \Delta}$ while
the elements of the affine deformation tensor (strains and local rotation), ${\mathsf
D}_{\alpha\gamma}$, arranged as a linear array ${\bf e} =
(D_{11}, D_{12}, \dots , D_{1d}, D_{21}, \dots , D_{dd})$, are given by ${\bf e} = {\mathsf Q}{\bf \Delta}$. Here 
${\bf \Delta}$ is a column vector with $Nd$ elements containing  the components of {\color{black} the} ${\bf \Delta}_i$.
The projectors are given explicitly by
${\mathsf R}{\mathsf Q}
= {\mathsf R}({\mathsf R}^{\rm T}{\mathsf R})^{-1}{\mathsf R}^{\rm T}$
and
${\mathsf P} = {\mathsf I}-{\mathsf R}{\mathsf Q}$.
As in~\cite{sassy}, the $Nd\times d^{2}$ matrix ${\mathsf R}$ appearing here has elements ${\mathsf
R}_{i\alpha,\gamma\gamma^{\prime}} =
\delta_{\alpha\gamma}(R_{i\gamma^{\prime}}-R_{0\gamma^{\prime}})$
where the $R_{i\gamma^{\prime}}$ and $R_{0,\gamma^{\prime}}$ are the
components of the lattice positions ${\bf R}_{i}$ and ${\bf R}_{0}$,
respectively. Now define the correlation matrix $\mathsf C$ with elements, $C_{i\alpha,j\gamma} = \langle \Delta_{i\alpha} \Delta_{j\gamma} \rangle$ where the angular brackets $\langle \dots \rangle$ indicate an average over the equilibrium ensemble. One can {\color{black}then} easily obtain the statistics of $\chi$ and ${\bf e}$ in terms of $\mathsf C$. For example the probability distribution for the affine distortions ${\bf e}$ is a $d^2$ 
dimensional Gaussian with zero mean and co-variance matrix ${\mathsf Q}{\mathsf C}{\mathsf Q}^{\rm T}$ whose elements are proportional to the elastic moduli. On the other hand, $\chi$ is distributed as the sum of the squares of $N_\Omega d-d^2$ independent Gaussian random variables with variances given by the eigenvalues of ${\mathsf P}{\mathsf C}{\mathsf P}$.  A comparison of the projected atomic displacements, i.e.\ eigenvectors of ${\mathsf P}{\mathsf C}{\mathsf P}$ and $(1-{\mathsf P}){\mathsf C} (1 - {\mathsf P})$, that give rise to the $\chi$ and $\bf e$ shows that while the latter consist of local volume, uniaxial and shear distortions of $\Omega$ together with local rotations, non-affine displacements, which contribute to $\chi$, correspond to small wavelength distortions of particles {\it within} $\Omega$.  Application of an external stress, ${\bf \Sigma}$, shifts the strain probability distributions to non-zero mean strain in accordance with Hooke's law and fluctuation response relations but does not affect $\chi$ to linear order. The lowest order variation of $\chi$ with ${\bf \Sigma}$ is given by $
 \langle\chi\rangle_{{\bf \Sigma}}=\langle\chi\rangle_{{\bf \Sigma}=0} + {\bf \Sigma}^{\rm T}{\mathsf Q}{\mathsf C}[{\mathsf P},{\mathsf C}]{\mathsf Q}^{\rm T}{\bf \Sigma}$, where $[{\mathsf P},{\mathsf C}]$ is a commutator.
 
In order to calculate the spatio-temporal correlation functions of the non-affinity $\chi$ and strains
${\bf e}$, we need to consider simultaneously 
displacement differences in two neighborhoods $\Omega$ and $\bar{\Omega}$ centered on
lattice positions $\bR_0$ and $\bar{\bR}_0$ at time $t$ and $t^{\prime}$ respectively.
 The vector
${\bf \Delta}(t)$ is defined as the displacement corresponding to the reference lattice position $\bR_0$ at time $t$, with an analogous definition for
$\bar {\bf \Delta}(t^{\prime})$.  The local
affine strain ${\bf e}(\bR_0,t) = {\mathsf Q} {\bf \Delta}(t)$ and non-affinity
$\chi(\bR_0,t) = {\bf \Delta}^{\rm T} (t){\mathsf P}{\bf \Delta}(t)$  are defined as before. For time $t^{\prime}$ and position $\bar{\bR}_0$ we have the corresponding quantities ${\bf e}(\bar{\bR}_0,t^{\prime}) = {\mathsf Q}\bar {\bf \Delta}(t^\prime)$ and $\chi(\bar{\bR}_0,t^{\prime}) = \bar{\bf \Delta}^{\rm T} (t^\prime){\mathsf P}\bar{\bf \Delta}(t^\prime)$. The covariances may now be defined as
\begin{eqnarray}
C_{i\alpha,j\gamma}&=&\langle \Delta_{i\alpha}(t)\Delta_{j\gamma}(t)\rangle
= \langle \Delta_{i\alpha}(0)\Delta_{j\gamma}(0)\rangle,  \nonumber \\
\bar{\bar{C}}_{i\alpha,j\gamma}&=&\langle \bar \Delta(t^{\prime})_{i\alpha}
\bar \Delta(t^{\prime})_{j\gamma}\rangle
= \langle \bar \Delta(0)_{i\alpha}
\bar \Delta(0)_{j\gamma}\rangle\nonumber \\
\bar {C}_{i\alpha,j\gamma}&=&\langle \Delta(t)_{i\alpha}
\bar \Delta(t^{\prime})_{j\gamma}\rangle.
\label{cmat}
\end{eqnarray}
Obviously the first two averages are identical and reduce to the space and time independent second-order moments $\langle {\bf \Delta}{\bf \Delta}^{\rm T} \rangle$~\cite{sassy}; the third quantity yields the required correlation functions.
To derive the expressions for the time-dependent strain and non-affinity auto-correlation functions we use their definitions in terms of the relative displacement projections. We obtain, therefore, 
\begin{eqnarray}
C_{\bf e}(\bR_0,t,\bar{\bR}_0,t^{\prime}) & = & \langle {\bf e}(\bR_0,t) {\bf e}^{\rm T}(\bar{\bR}_0,t^{\prime})\rangle \nonumber \\
& = &\langle {\mathsf Q} {\bf \Delta}(t) \bar {\bf \Delta}^{\rm T}(t^{\prime}) {\mathsf Q}^{\rm T} \rangle \nonumber \\
& = & {\mathsf Q} \bar {\mathsf C} {\mathsf Q}^{\rm T}.
\label{ep}
\end{eqnarray}
The correlation functions between any pair of affine {\color{black}strains} may now be obtained by taking appropriate linear combinations of the elements of  $C_{\bf e}$. In the next section we focus on one such component, viz, the shear strain $\epsilon$.
Similarly, the correlation between $\chi(\bR_0,t)$ and $\chi(\bar{\bR}_0,t^{\prime})$ can be calculated using Wick's theorem as 
\begin{eqnarray}
C_{\chi}(\bR_0,t,\bar{\bR}_0,t^{\prime}) & = & \langle \chi(\bR_0,t)\chi(\bar{\bR}_0,t^{\prime})\rangle -  \langle\chi\rangle^2\nonumber\\
& = &2\,{\rm Tr} ({\mathsf P}\bar {\mathsf C}{\mathsf P})
({\mathsf P}\bar{\mathsf C}{\mathsf P})^{\rm T} = 2\sum_j\bar{\sigma}_{j}^{2}
\label{chi}
\end{eqnarray}
where, in the final equation, the $\bar{\sigma}_{j}^2$
denote the $N_\Omega\,d - d^2$ non-zero eigenvalues of the matrix $({\mathsf P}\bar{\mathsf C}{\mathsf P})
({\mathsf P}\bar{\mathsf C}{\mathsf P})^{\rm T}$. 
Of course, in a homogeneous solid in equilibrium, these correlation functions are functions only of the relative coordinates ${\bf R}_0 - \bar{\bf R}_0$ and times $t - t^{\prime}$. We will denote these simply by ${\bf R}$ and $t$ in what follows.

Note that so far we have not made any assumptions about the structure and interactions of the particles $i$ and all our results apply equally well for any system in any dimension as long as a well defined reference configuration $\{ {\bf R} \}$ exists. Indeed, we believe that a fair fraction of our results should apply even to {\em amorphous} solids with displacements being measured from a set of particle coordinates obtained from a zero temperature energy minimisation. To obtain analytic results we need to evaluate the covariances and for the rest of this paper we specialise to periodic lattices of particles, whose interactions we may approximate as being {\em harmonic}. Alternately, the covariance matrix may also be obtained experimentally~\cite{harm-colloid} in the case of colloidal solids using video microscopy without any {\em a priori} assumption concerning the form of the interactions. One may directly measure 
$\langle {\bf u}_{\bf q} {\bf u}_{-{\bf q'}}^{\rm T} \rangle = \Dmat^{-1}({\bf q}) \,\vBZ\delta({\bf q}-{\bf q'})$, where $\Dmat({\bf q})$ is the dynamical matrix, and $\vBZ$ the volume of the Brillouin zone. Given the dynamical matrix, $\bar{C}_{i\alpha,j\gamma}$ may be evaluated as follows. We substitute
for the relative displacements their expansion in terms of the vibrational modes of the lattice viz., ${\bf \Delta}_i (t) = {\bf u}_i - {\bf u}_0 = l\, \vBZ^{-1} \sum_s \int d{\bf q}\, {\bf u}_{\bf q}^{\rm T} {\bf a}_s({\bf q}){\bf a}_s({\bf q}) (e^{i {\bf q \cdot  R}_{i}} - e^{i {\bf q \cdot R}_0})\cos[\omega_{s}({\bf q})t] $, {\color{black}into} the third of the equations~(\ref{cmat}) to obtain,
\begin{eqnarray}
 \bar{C}_{i\alpha,j\gamma}&=&\frac{l^{2}}{v_{BZ}}\sum_s\int\,d{\bf q}\, {\color{black}\,a_{s\alpha}}({\bf q}) {\color{black}\,a_{s\gamma}}({\bf q})\frac{\cos[\omega_{s}({\bf q})t]}{\omega^{2}_s({\bf q})} \times \nonumber \\
 & & (e^{i {\bf q \cdot R}_{i}}-e^{i {\bf q \cdot R}_{0}}) (e^{-i {\bf q \cdot \bar{R}}_j}-e^{-i {\bf q \cdot \bar{R}}_{0}}).
\label{MATC}
 \end{eqnarray}
In the above expressions $l$ is the lattice parameter and ${\color{black}{\bf a}}_{s}({\bf q})$ and $\omega_{s}({\bf q})$ are the eigenvectors and eigenvalues (phonon frequencies) respectively of the dynamical matrix corresponding to the $s^{\rm th}$ phonon branch. The ${\bf q}$--space integrals are over the Brillouin zone. 

\section{Results for the 2d triangular crystal}
\label{sec:2}

The formulation for the spatio-temporal correlation functions given in the previous section (section~\ref{sec:1}) is applicable for any periodic crystal as long as the dynamical matrix $\Dmat_{\alpha\gamma}$ is known. In this section we present our results for the simple but important case of a triangular network of particles connected by harmonic springs defined by the Hamiltonian, 
\begin{equation} 
H_{harm} = \sum_{i} \frac{{\bf p}_i^2}{2 m} + \frac{K}{2} \sum_{(ij)} [({\bf u}_i - {\bf u}_j)\cdot {\hat {\bf R}_{ij}}]^2,
\label{harm}
\end{equation}
where ${\bf u}_i$, ${\bf p}_i$ and  $m$ are displacement, {\color{black}momentum} and mass of the particle $i$ respectively, and the unit vector ${\hat {\bf R}_{ij}}$ points from particle $i$ to $j$ in the reference configuration. The sum in the second term in (\ref{harm}) runs over all bonds in the network, each with spring constant $K$. The unit of distance will be the lattice parameter from now on while time will be measured in units of $\sqrt{m/K}$. The temperature may also be rescaled to unity without loss of generality. Because of its simplicity, the harmonic triangular net has been studied extensively and is known to be a good approximation for many real crystalline solids in two dimensions~\cite{zahn,kers1,zhang,harm-colloid}. The dynamical matrix and hence the dispersion relation $\omega({\bf q})$ for this system is also known~\cite{harmdyn}; for small ${\bf q}$ it is given by $\omega_{s = T,L} = c_{s = T, L} |{\bf q}|$ with the transverse and longitudinal sound velocities $c_T = \frac{1}{2}\sqrt{\frac{3 K}{2}}$ and $c_L = \frac{3}{2}\sqrt{\frac{K}{2}}$. We consider a coarse-graining volume $\Omega$ consisting of a central atom and its $N_\Omega = 6$ nearest neighbours in the triangular lattice~\cite{sassy}. 
 \begin{figure}[ht]
\begin{center}
\includegraphics[width = 6.0cm]{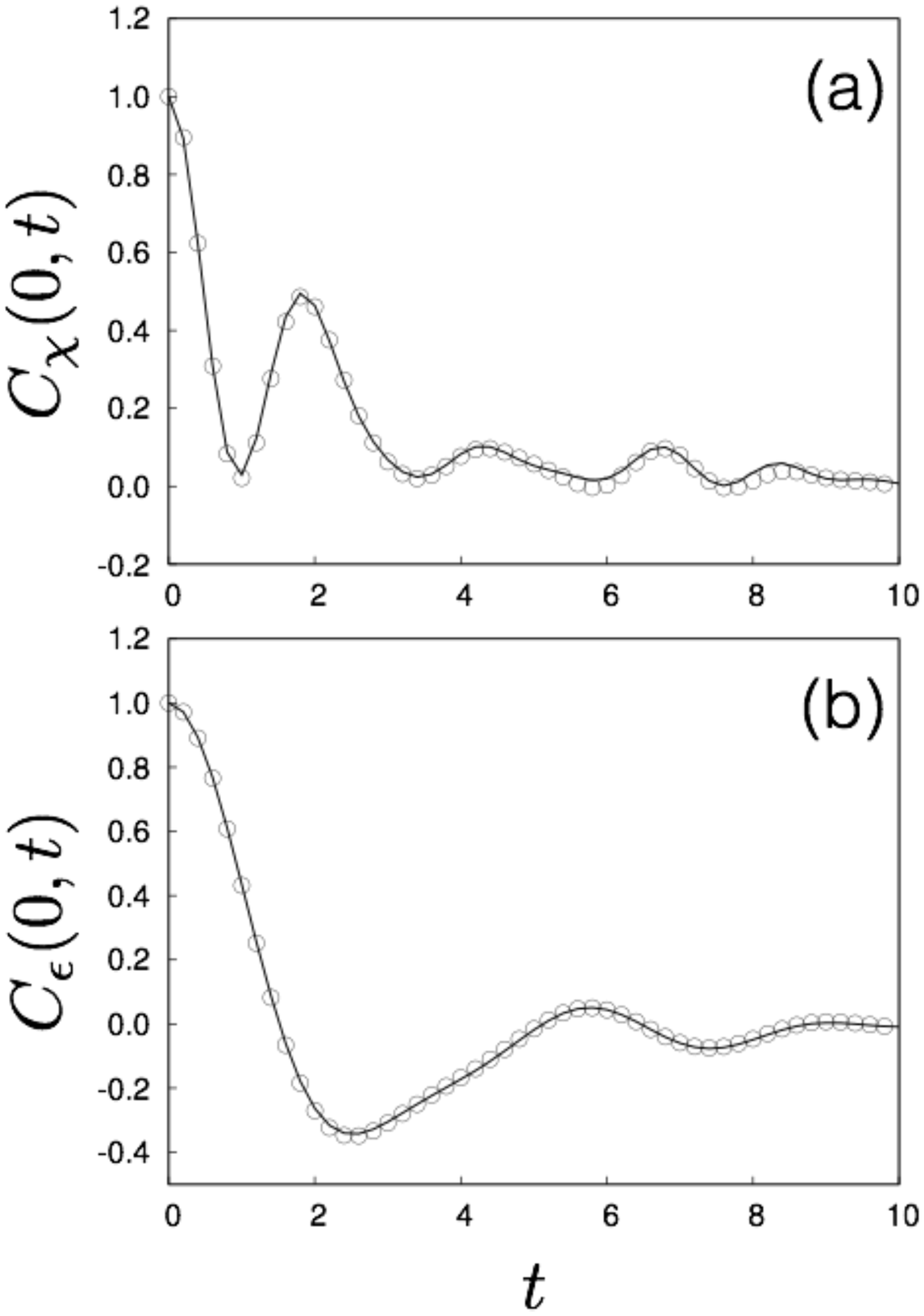}
\end{center}
\caption{(a) $C_{\chi}(0,t)$ from computation (line) compared with
  that obtained from molecular dynamics simulations (points) of a
  $500\times500$ site harmonic lattice with unit particle masses. Note that the error bars are smaller than the size of the symbols.  (b) Plot of $C_{\epsilon}(0,t)$ for the same system as in (a). 
}
\label{corrt}
\end{figure}
The normalised 
correlation functions for $\chi$ and $\epsilon$ fluctuations are given by $$C_{\chi}({\bf R},t) = \frac{\langle\chi(0,0)\chi({\bf R},t)\rangle-\langle\chi\rangle^2}{\langle\chi^2\rangle-\langle\chi\rangle^2}$$ and $$C_{\epsilon}({\bf R},t)= \frac{\langle \epsilon(0,0)\epsilon({\bf R},t)\rangle}{\langle \epsilon^2\rangle},$$ where $\epsilon = (D_{12}+D_{21})/2 = {\bf e}_3$ is the shear strain. Note that the temperature (or coupling constant) scales out for these quantities. 

In Fig.\ref{corrt}(a), we plot $C_{\chi}({\bf 0},t)$. 
The integrals over the Brillouin zone were computed numerically using a $256$ point Gauss quadrature for each real space lattice position. Our results are compared with those obtained from molecular dynamics simulations of a $500\times500$ site harmonic triangular net.  To obtain the simulation results, the solid was first allowed to equilibrate for $4 \times 10^5$  MD steps with a time-step of $5\times10^{-4}$ in the canonical (NVE) ensemble. 
Data was collected from $4 \times 10^5 $ to $4.2 \times 10^5$ MD steps at intervals of $400$ MD steps. Correlation functions were obtained by averaging over particles.  

It is clear that our results agree with simulation data within the error bars of the latter. The decay of $C_\chi$ is not monotonic but oscillatory, a feature arising from the time-periodic lattice vibrations of the solid, which are all in phase at $t=0$, gradually de-cohering for larger times~\cite{vine}.  Similar oscillations are also observed in $C_{\epsilon}$ shown in Fig.\ref{corrt}(b) for the same system and both $\chi$ and $\epsilon$ relax over similar time scales. 

\begin{figure}[h]
\begin{center}
\includegraphics[width=6.0cm]{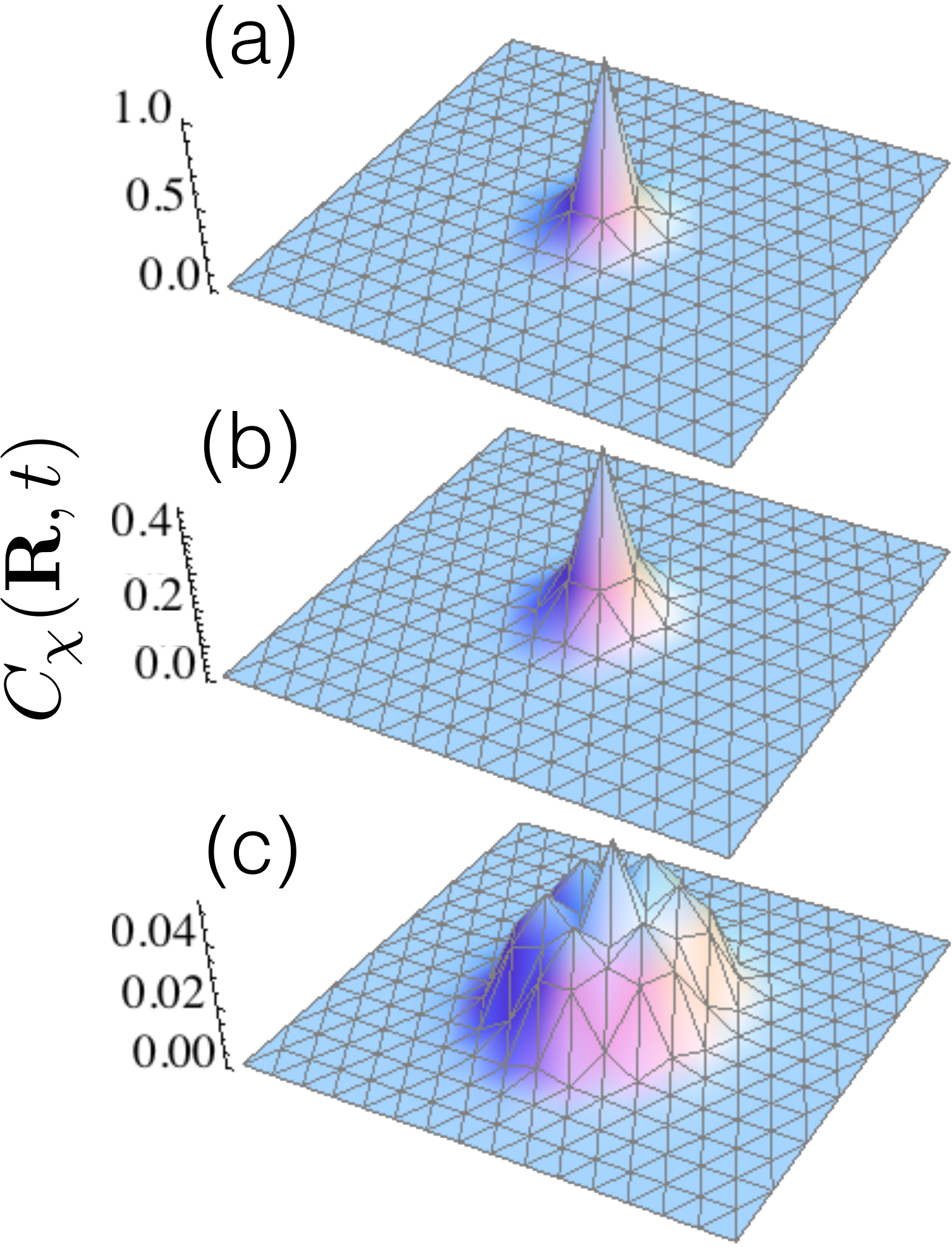}
\caption{$C_\chi({\bf R},t)$ calculated at different times (a) $t=0$, (b) $t=2$, (c) $t=5$ for the harmonic triangular lattice. The parameters are the same as in Fig.\ref{corrt}. Note changes in scale between the three subpanels.}
\label{CHICor}
\end{center}
\end{figure}

\begin{figure}[h]
\begin{center}
\includegraphics[width=6.0cm]{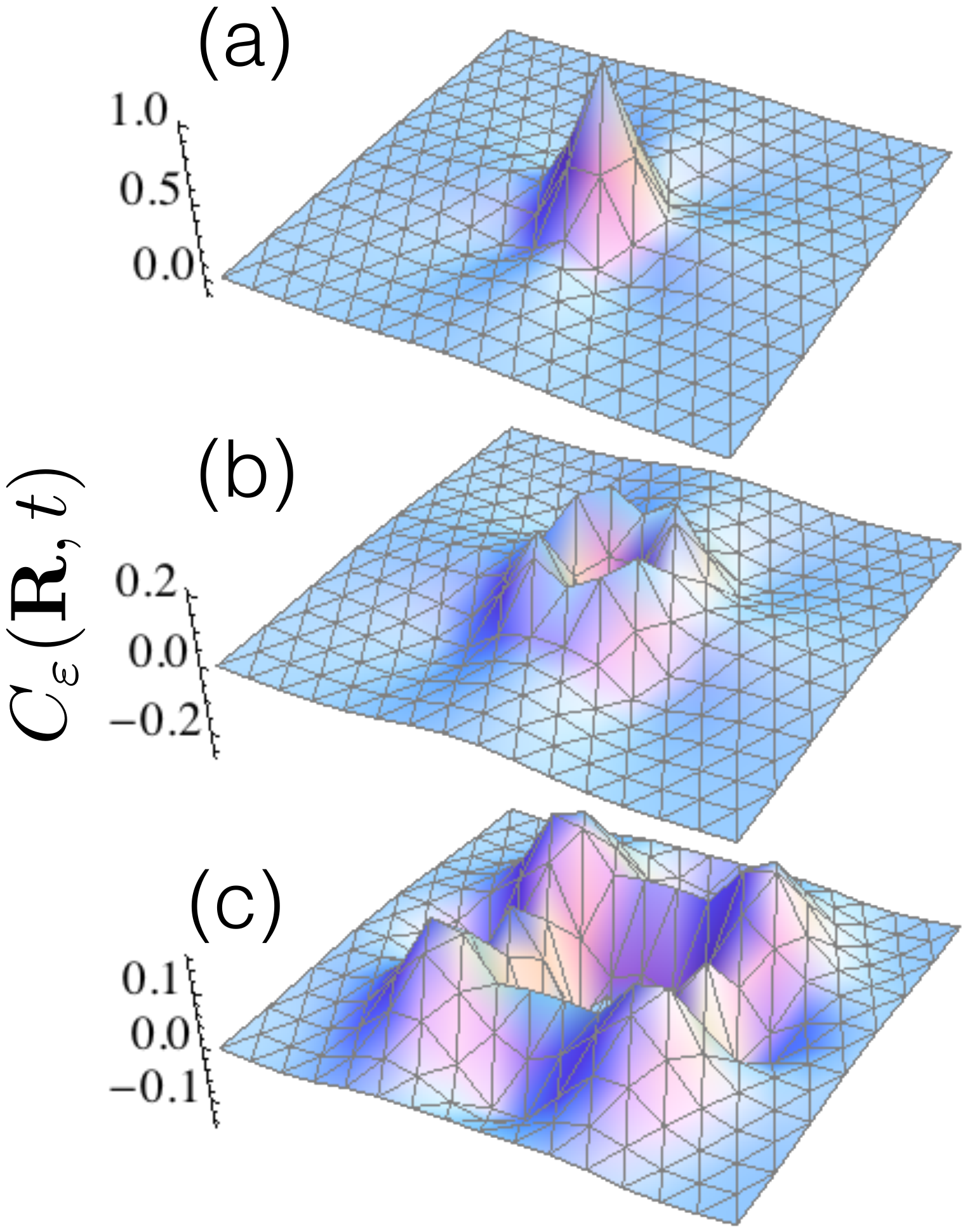}
\caption{$C_\varepsilon({\bf R},t)$ calculated for times (a)  $t=0$, (b) $t=2$, (c) $t=5$ for the same set of parameters as in Fig.\,\ref{corrt}. Note changes in scale between subpanels.}
\label{EPCor}
\end{center}
\end{figure}

The complex relaxation of the dynamic correlations is even more in evidence when we evaluate these functions in both space and time. In Fig.\ref{CHICor}(a)-(c) we plot the full $C_\chi({\bf R},t)$ for the first few nearest neighbour lattice points of the 2d triangular lattice. For small times, this {\color{black} correlation} function is sharply peaked at the origin and decays rapidly to zero after the second neighbour shell (Fig.\ref{CHICor}(a)). At larger times, the function decays but becomes longer ranged extending up to the sixth neighbour shell for $t=5$ (Fig.\ref{CHICor}(c)), corresponding to a spread with a speed comparable to $c_T$;  finally, $C_\chi({\bf R},t) \to 0$ everywhere for large time differences $t$.
In Fig. \ref{EPCor}(a)-(c) we show the corresponding space-time correlation functions for the shear strain. The equal time spatial correlation Fig.\ref{EPCor}(a) has been calculated before~\cite{sassy} and has also been measured from video microscopy of colloidal solids\cite{kers1}. The typical four-fold symmetry (butterfly pattern) of this correlation function has also been observed in experiments on amorphous colloids \cite{schall}. This pattern is easy to understand since $C_{\epsilon}({\bf R},0)$ represents the response of a solid to a delta function shear load at the origin, which may arise from a small inclusion or ``Eshelby'' defect ~\cite{quad}. At subsequent times, the correlation function retains its significant four-fold symmetry, although it shows wave-like oscillations in space and time; these eventually decay to zero.
 
The four-fold symmetry of the strain correlation function can also be understood from a momentum space formulation as described in~\cite{sassy}, where the Fourier transform of the shear strain correlation function is simply given by
$$ \tilde C_{\epsilon}({\bf q},t) = {\mathsf E}_{1212}+{\mathsf E}_{2121} + 2 {\mathsf E}_{1221},$$ with
\begin{equation}
{\color{black}{\mathsf E}_{\alpha \alpha' \gamma \gamma'} ({\bf q},t) = q_{\alpha'} q_{\gamma'}\sum_{s}\frac{{\color{black}\,a}_{s\alpha}({\bf q}) {\color{black}\,a}_{s\gamma}({\bf q})}{\omega^{2}_s({\bf q})} \cos(\omega_{s}({\bf q})t)}
\end{equation}
Substituting the expressions for ${\mathsf E}$ into $\tilde C_{\epsilon}$ and expanding all ${\bf q}$ dependent quantities to leading order in the wavenumber, we finally get ($q=|{\bf q}|$)
\begin{eqnarray}
\tilde C_{\epsilon}({\bf q},t)& = &\frac{4q_{x}^{2}q_{y}^{2}}{c_{L}^{2}{\bf q}^4}\cos(c_{L}\,q\,t)+\frac{(q_{x}^{2}-q_{y}^{2})^{2}}{c_{T}^{2}{\bf q}^4}\cos(c_{T}\,q\,t).\nonumber
\end{eqnarray}
The four-fold symmetry of $\tilde C_{\epsilon}$ is now obvious.

\section{Defect precursors in the 2d triangular crystal}
\label{sec:3}
In the previous two sections we derived a systematic procedure for analysing particle displacements within a coarse-graining volume $\Omega$ as affine or non-affine. The affine displacements can be identified as elastic strains whose fluctuations determine the elastic constants of the solid. In this section we turn to the identity of the non-affine fluctuations and show that {\color{black}the fluctuations with the highest contribution to $\chi$ represent precursors to the formation of pairs of lattice defects.}  We also show that they are statistical fluctuations that obey standard fluctuation-response relations and thereby identify the conjugate field {\color {black}$h_X$}. Positive values of {\color {black}$h_X$} enhance and negative values suppress lattice defects. Finally we calculate space-time correlation functions for these lattice distortions in the presence of nonzero {\color {black}$h_X$}, using results derived in the earlier sections. 

\begin{figure}[t]
\begin{center}
\includegraphics[width=6.0cm]{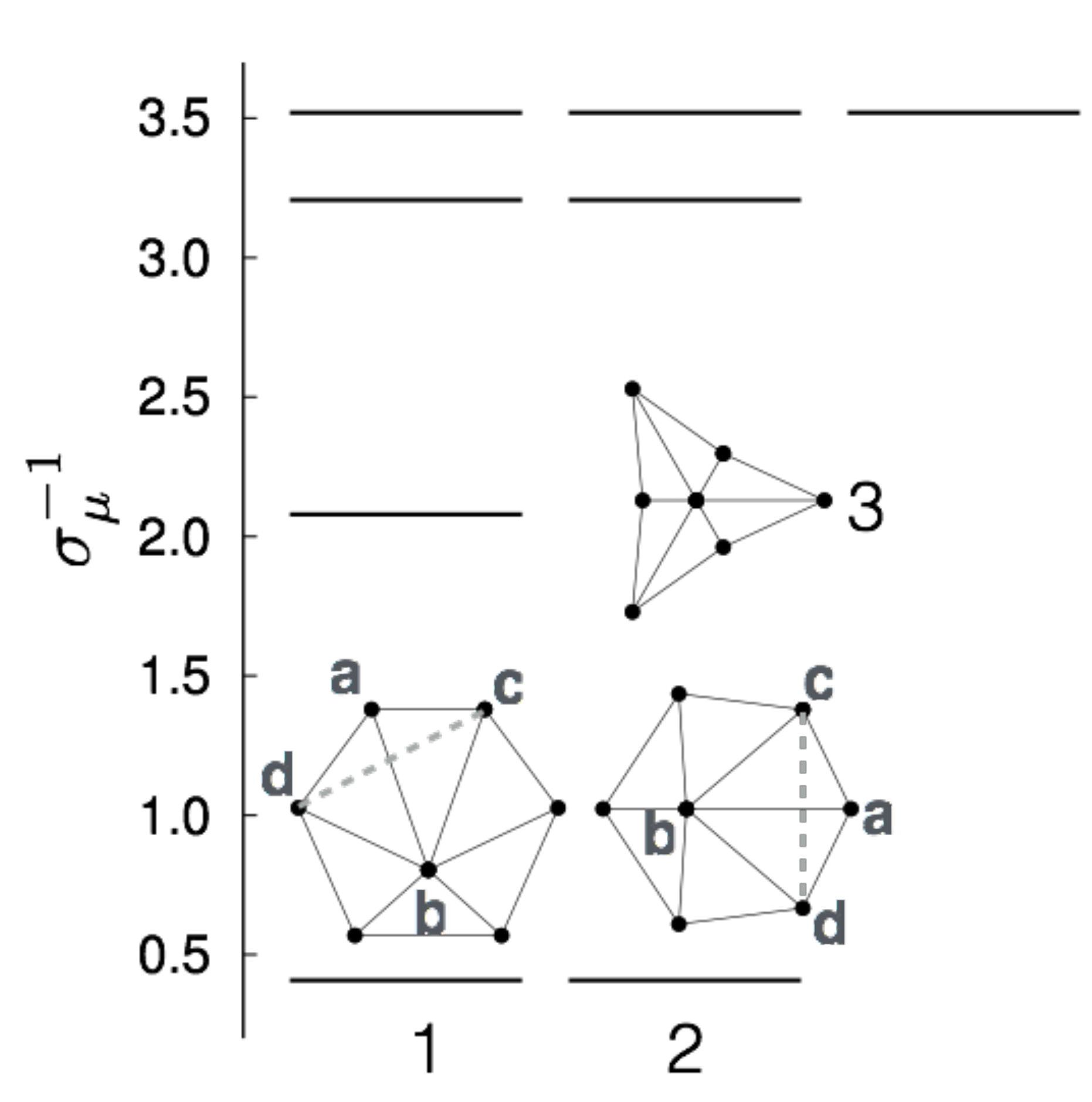}
\caption{Plot of the inverses of the eigenvalues 
$\sigma_\mu$ of the non-affine projection ${\mathsf P}{\mathsf C}{\mathsf P}$ together with the eigenvectors for the three largest eigenvalues that indicate the most prevalent fluctuations. Note that there are two degenerate soft modes $1,2$ separated by a large gap from the next most important contributor $3$. The two soft modes tend to increase the nearest neighbour bond length $a-b$, at the same time decreasing the distance between next nearest neighbours $c-d$ (shown by gray dashed line).}
\label{eigen}
\end{center}
\end{figure}

Recall that in the 2d triangular lattice $\langle \chi \rangle = \sum^8_{\mu=1} \sigma_\mu$ where the $\sigma_\mu$ are the eight non-zero eigenvalues of the $N_\Omega\,d \times N_\Omega\,d =12\times12$-dimensional matrix ${\mathsf P}{\mathsf C}{\mathsf P}$~\cite{sassy}. The eigenvectors ${\color{black}{\bf b}_{\mu}}$ corresponding to these eigenvalues represent non-affine distortions of the coarse graining volume, their relative contributions to $\chi$ being determined by the value of $\sigma_\mu$. In Fig.~\ref{eigen} we plot the magnitudes of $\sigma_\mu^{-1}$. 
It is immediately clear that there are three groups of terms. {\color{black}The eigenvalues of the two degenerate, non-affine modes corresponding to $\mu = 1$ and $\mu = 2$ are separated from the next higher one $\mu = 3$ by a large gap -- a factor of $4$ -- and from the rest by an order of magnitude.} A close look at the eigenvectors corresponding to these eigenvalues reveals that these non-affine distortions tend to increase the distance between nearest neighbour particles and {\em reduce} next nearest neighbour bond lengths. If a nearest-neighbour bond is actually replaced by a next-nearest-neighbour one, then the coordination number of the particles changes and a pair of particles with $5$ and $7$ neighbours each would emerge out of the reference $6$-coordinated triangular structure. Each pair of neighbouring $5$- and $7$-coordinated atoms contains a dislocation (or an anti-dislocation depending on the orientation). These dislocation-anti-dislocation pairs can then separate from each other by subsequent non-affine fluctuations that change the coordination number of neighbouring atoms. 

Of course in a harmonic lattice defects do not nucleate, though non-affine precursor fluctuations exist. Indeed, the overlap of particle displacements with a non-affine eigenvector ${\color{black}{\bf b}_{\mu}}$, given by $s_\mu = {\color{black}{\bf b}_{\mu}}^{\rm T}{\bf \Delta}$ 
, is a Gaussian random variable with probability distribution, 
\begin{align*}
P(s_\mu)&=\frac{1}{\sqrt{2\pi{\color{black}{\bf b}_{\mu}}^{\rm T}\mathsf{C}{\color{black}{\bf b}_{\mu}}}}\exp\left(-\frac{s_\mu^{2}}{{\color{black}{\bf b}_{\mu}}^{\rm T}\mathsf{C}{\color{black}{\bf b}_{\mu}}}\right),
\end{align*}
an expression analogous to the one for strains. The quantity ${\color{black}{\bf b}_{\mu}}^{\rm T}\mathsf{C}{\color{black}{\bf b}_{\mu}}$ appears as a susceptibility for defect precursor fluctuations. The fluctuation-response relation connects this susceptibility with a response function measuring the response of $s_\mu$ to a conjugate field. We investigate this connection below. For the rest of this paper, we present results only for $\mu = 1$. The corresponding {\color {black} results} for the degenerate $\mu = 2$ eigenvector are either identical or completely analogous.  

To proceed further, we consider the global non-affinity $X = N^{-1}\sum_{i=1}^N \chi{\color{black} ({\bf R}_i)}$ averaged over all particles $i = 1\dots N$ 
 and introduce a field conjugate to this quantity in (\ref{harm}) to obtain,
\begin{equation}
H = H_{harm}  - h_X\,N\,X.
\label{harmh}
\end{equation}
The extra term in (\ref{harmh}), though still quadratic in the particle coordinates, introduces a many-body force (see section~\ref{sec:4}) that depends on the positions of all particles in a given neighbourhood {\color{black}$\Omega$}. A change in the coordinate of particle $i$ modifies not only the local {\color{black}$\chi$} at that particle but also those of its neighbours. The force also depends on the reference lattice positions $\{{\bf R}\}$, which act as constant parameters. A purely affine transformation of $\Omega$, such as a volume rescaling for example, does not produce a non-affine force. This force therefore tracks only non-affine distortions away from the reference configuration. The dynamical matrix corresponding to (\ref{harmh}) can be computed without difficulty and therefore the statistics of the local $\chi$ and the local strains, 
together with their space-time correlation functions can be obtained for arbitrary $h_X$ using the procedures outlined in~\cite{sassy} and sections~\ref{sec:2} and~\ref{sec:3}. This holds true as long as the structure of the solid is maintained, i.e.\ as long as the reference configurations $\{{\bf R}\}$ remain the global minimum of the modified Hamiltonian (\ref{harmh}). Note that, for our MD simulations, in addition to the term proportional to $h_X$, we have also included a small hard core repulsion of the Weeks,Chandler, Anderson (WCA) form~\cite{UMS}. This prevents atoms from overlapping at large values of $h_X$ and also introduces anharmonicity in a controlled fashion such that {\color{black} the} relative contribution of this term to the energy provides us with a measure of anharmonic contributions. The hard core diameter $d_0 = 0.6 l$ was chosen to be small enough so that anharmonic effects vanish for {\color{black}small values of $h_X$} and all our results based on harmonic analysis hold in this limit. 

The statistics of the global $X$ and the local $\chi$ are clearly related to each other. For example, the thermal averages are equal, $\langle X \rangle = \langle \chi \rangle$. The variance of $X$ is 
\begin{eqnarray}
\langle (\Delta X)^2 \rangle & = & \langle X^2\rangle  - \langle X \rangle^2 \nonumber \\
& = & N^{-2} \langle \sum_{\bf R} \chi({\bf R}) \sum_{{\bf R}'} \chi({\bf R}') \rangle - \langle \chi \rangle^2 \nonumber \\ 
& = & N^{-1} \sum_{R}[\langle \chi({\bf 0})\chi({\bf R}) \rangle - \langle \chi \rangle^2] \nonumber \\
& = & N^{-1} \langle {\color{black}(\Delta \chi)^2} \rangle \sum_{\bf R}C_\chi({\bf R},0)
\label{therml}
\end{eqnarray} 
The cross-correlation $\langle \Delta\chi({\bf R}) \Delta X \rangle$ has the same expression for any $\bf R$. 

\begin{figure}[t]
\begin{center}
\includegraphics[width=6.0cm]{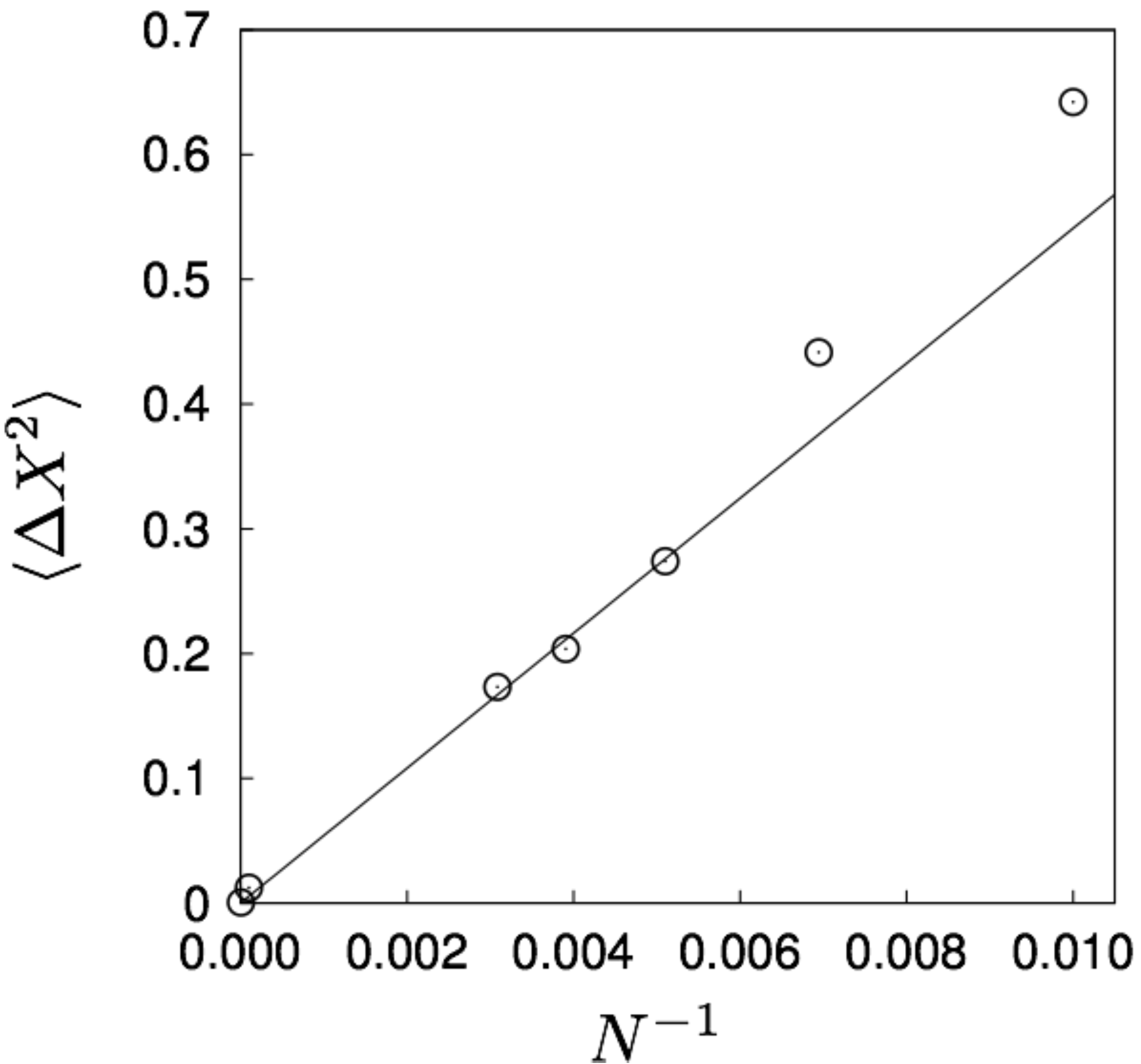}
\caption{Plot of $\langle (\Delta {\color {black} X})^2 \rangle$ as a function of $N^{-1}$ at ${\color {black}h_X} = 0$ from our MD simulations for $N = 10\times10, 12\times12, 14\times14, 16\times16, 18\times18, 100\times100$ and $500\times500$ lattices. The straight line is the prediction from (\ref{therml}) without any fitting parameters. Note that for smaller lattices there are significant deviations from the asymptotic slope.}
\label{therm-limit}
\end{center}
\end{figure}

The variance of $X$ vanishes in the $N\to \infty$ limit as expected for an intensive thermodynamic variable (see Fig.~\ref{therm-limit}); the distribution, $P(X)$, therefore becomes a delta function centered at $\langle \chi \rangle$.
To obtain the response $\langle X(h_X)\rangle$, for small $h_X$ we first compute $\langle {\color{black}(\Delta\chi)^2} \rangle$ at $h_X = 0$ and then use the linear response relation, 
\begin{equation}
\frac{\partial \langle X \rangle}{\partial h_X} = \langle {\color{black}(\Delta \chi)^2} \rangle \sum_{\bf R} C_\chi({\bf R},0).
\label{LR-chi}
\end{equation}
\begin{figure}[t]
\begin{center}
\includegraphics[width=6.0cm]{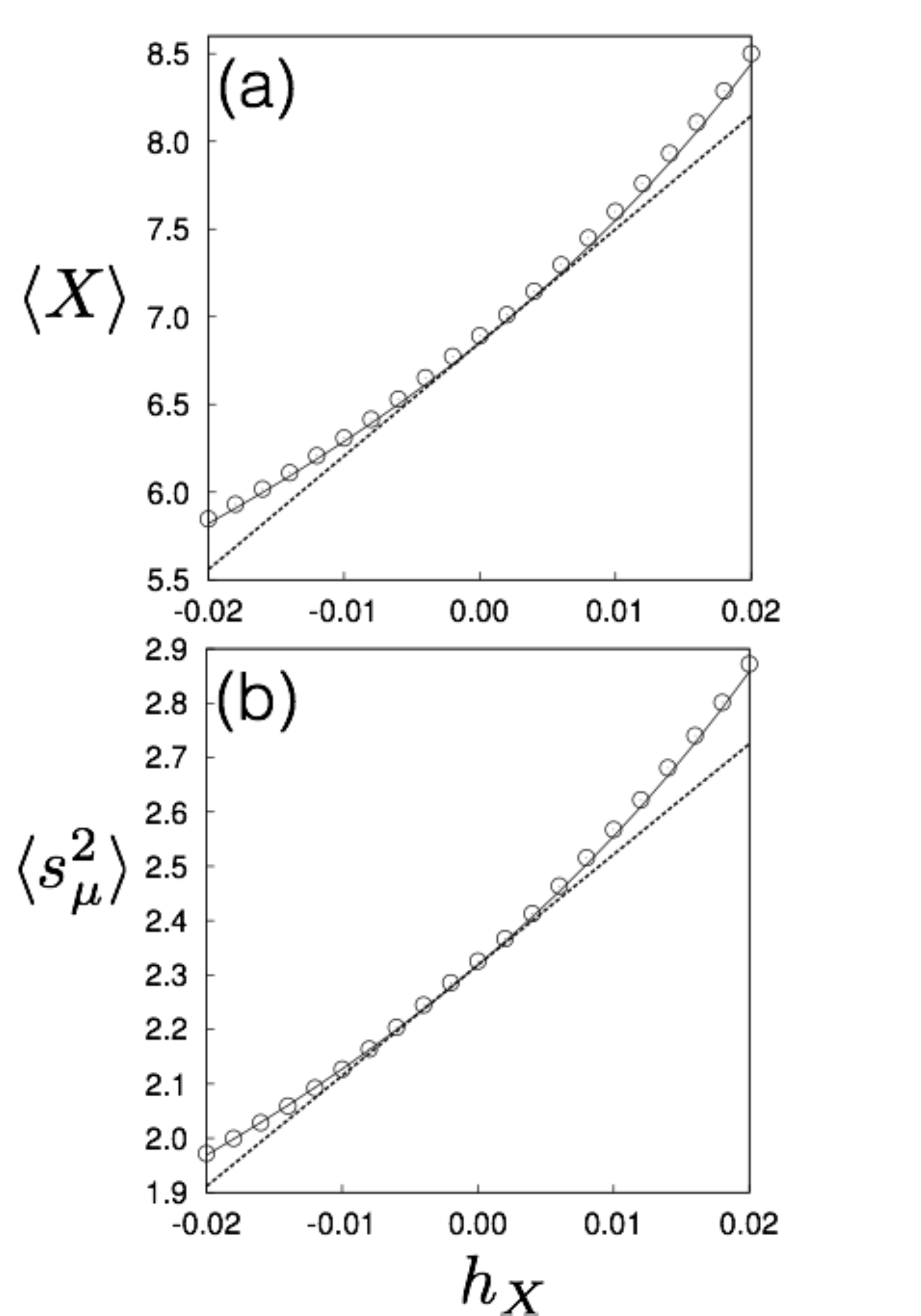}
\caption{Plots of {\color {black} (a)} $\langle {\color {black} X} \rangle$ and {\color {black}(b)} $\langle s_{\mu}^2 \rangle$, for $\mu = 1$, as a function of $h_X$. The points are MD simulation data from an $N = 200\times200$ lattice. Error bars are smaller than the size of the symbols. The solid curves through the data are the analytic results obtained from direct computation. The dashed straight lines are {\color{black}the} linear response predictions (\ref{LR-chi}) \& (\ref{LR-s}) respectively.} 
\label{resp}
\end{center}
\end{figure}

The non-affine field also changes the statistics of the {\color{black}dominant displacement fluctuations $s_\mu$}.
Since the global non-affinity $X$ is quadratic in particle displacements, the field $h_X$ cannot break the symmetry of $s_\mu$. The probability distribution $P(s_{\mu})$ remains Gaussian but with a variance $\langle s_{\mu}^2 \rangle$ that depends on {\color {black}$h_X$}. Again, a linear response calculation gives, 
 \begin{eqnarray}
\frac{\partial \langle  s_{\mu}^{2} \rangle}{\partial h_X} & = & \langle X s_{\mu}^{2}\rangle_{0}-\langle X\rangle_{0}\langle s_{\mu}^{2}\rangle_{0} \nonumber \\
& = & 2 \sum_{\bf R} {\color{black}{\bf b}^{\rm T}_{\mu}}\bar{\mathsf{C}} \mathsf{P} \bar{\mathsf{C}}^{\rm T} {\color{black}{\bf b}_{\mu}}.
\label{LR-s}
\end{eqnarray}
with $\bar{\mathsf{C}}$ as given in (\ref{MATC}).

In Fig.~\ref{resp} we have plotted $\langle {\color{black} X} \rangle$ and $\langle s_{\mu}^2 \rangle$ as functions of $h_X$. For $h_X > 0$, both of these quantities increase making defects more likely to form. In contrast a negative $h_X$ {\em suppresses} those fluctuations that give rise to defects. The points are simulation results that are compared with the linear response results as well as the full nonlinear calculation obtained by evaluating the dynamical matrix for the Hamiltonian~(\ref{harmh}).

\begin{figure*}[t]
\begin{center}
\includegraphics[width=15.0cm]{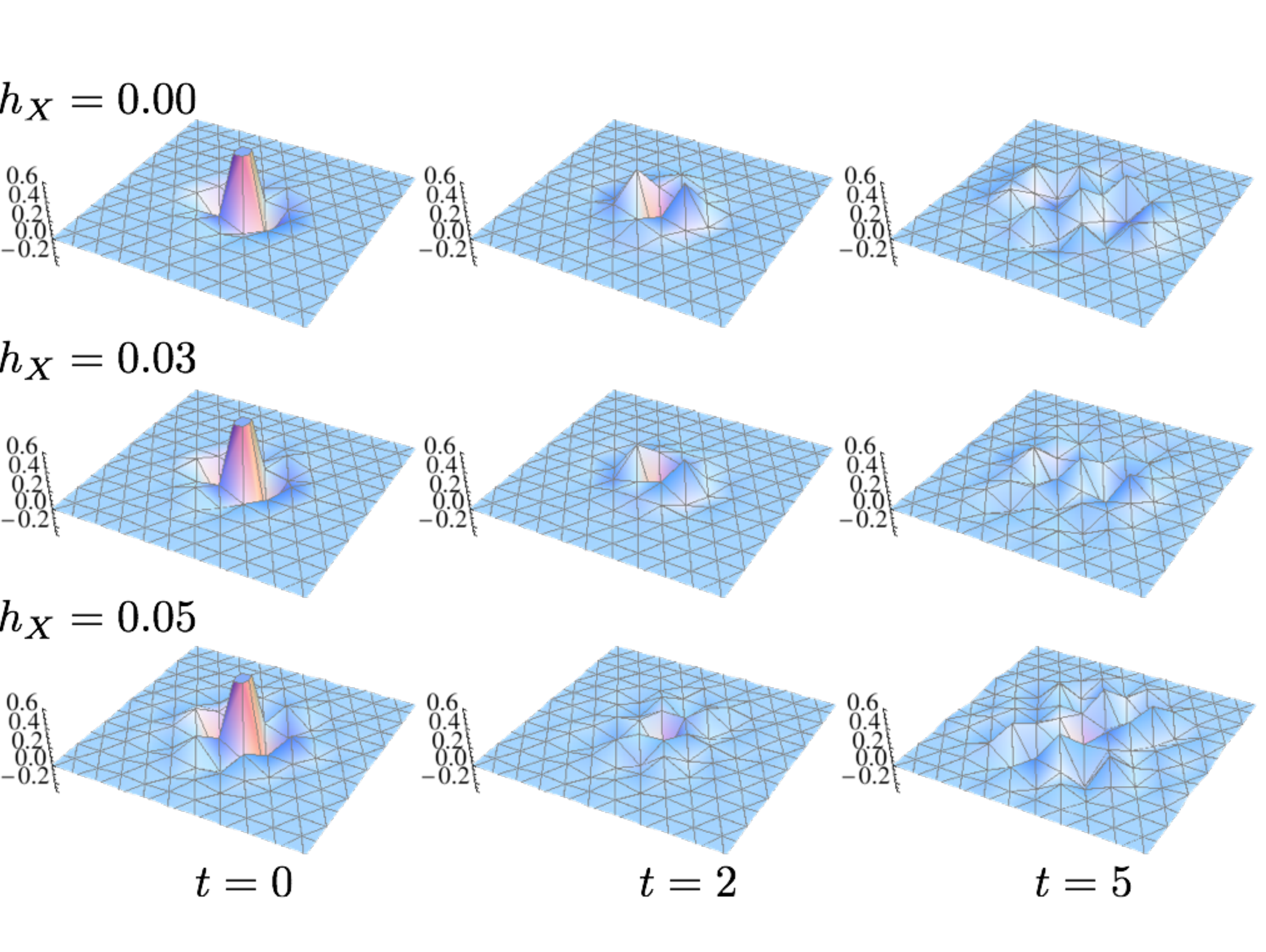}
\caption{Plot of the normalised  $C_s({\bf R},t)$ for ${\color {black}h_X} = 0.00, 0.03$ and $0.05$ for $t = 0$, $2$ and $5$. Note that the precursor fluctuations are spatially anisotropic and anti-correlated at all times. The non-affine field makes the correlations stronger. We have plotted the correlations for $\mu = 1$; the corresponding functions for $\mu = 2$ are essentially rotated by $90^\circ$.} 
\label{mu-rtc}
\end{center}
\end{figure*}

Space-time correlations of the defect precursors may be computed quite straightforwardly from the formalism presented in section~\ref{sec:2}. Indeed, the correlation of the dominant non-affine displacement $s_{\mu}$ is 
$\langle s_\mu^2 \rangle\,C_s({\bf R},t) = \langle s_{\mu}({\bf 0},0)s_{\mu}({\bf R},t)\rangle = {\color{black}{\bf b}_{\mu}}^{\rm T} \bar{\mathsf{C}}{\color{black}{\bf b}_{\mu}}.$ 
\begin{figure}[t]
\begin{center}
\includegraphics[width=6.0cm]{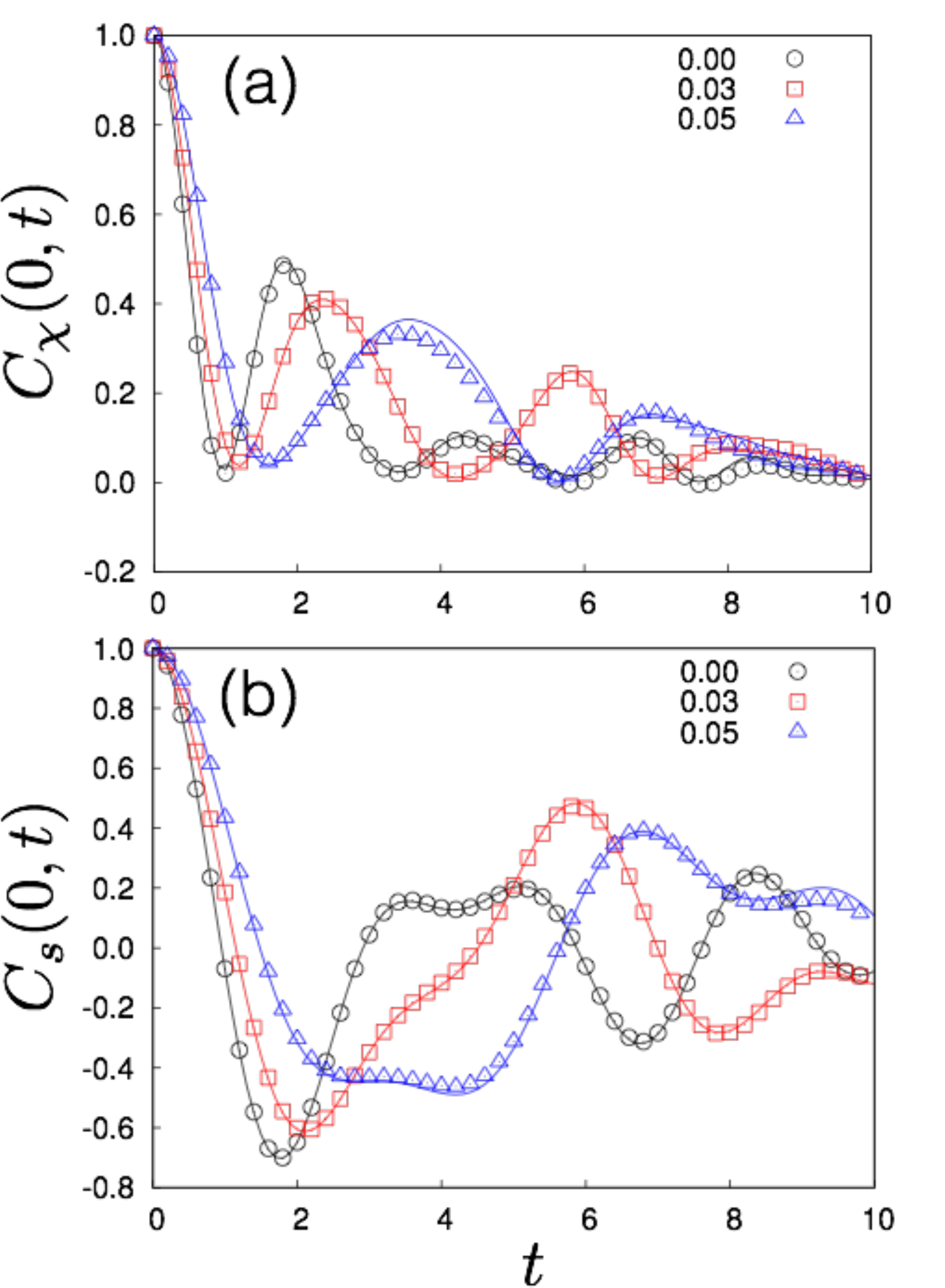}
\caption{ Plots of the normalised {\color{black}(a)} $C_\chi({\bf 0},t)$ and {\color {black}(b)} $C_s({\bf 0},t)$ for three values of $h_X$ compared with results from MD simulations of $500 \times 500$ particles. Note that the precursor fluctuations become more long-lived as {\color {black}$h_X$} increases.} 
\label{mu-tc}
\end{center}
\end{figure}
Fig.~\ref{mu-tc} shows plots of  $C_\chi({\bf 0},t)$ and $C_s({\bf 0},t)$ 
against time $t$ for a few values of {\color {black}$h_X$}. The displacement correlations, like those shown in section~\ref{sec:3}, are oscillatory and decay slowly in time due to destructive interference of the large number of mutually incommensurate phonon modes that make up these localised fluctuations. More importantly, Fig.~\ref{mu-tc} shows that the lifetime of these defect precursors grows as {\color {black}$h_X$} increases; the time period of the correlation function oscillations also increases as expected.  Finally in Fig.~\ref{mu-rtc} we plot the full $C_s ({\bf R},t)$ for three values of ${\color {black}h_X} = 0.00, 0.03$ and $0.05$ as well as for three values of the time $t=0, 2$ and $5$ as in Figs. ~\ref{CHICor} and \ref{EPCor}. Unlike the correlation functions for $\chi$ and $\varepsilon$, the correlations of $s_{\mu}$ are ``anti-ferromagnetic", i.e. a fluctuation $s_{\mu}$ of any sign at some lattice point induces a fluctuation of $s_{\mu}$ of the {\em opposite} sign at the neighbouring lattice point. 

What is the effect of external stress $\bf{\Sigma}$ on the defect precursors? It is again easy to answer this question by a straightforward calculation: one only needs to include the term {\color{black}${\bf \Sigma}^{\rm T} \sum_{i=1}^{N} {\bf e}({\bf R}_{i})$}
 in the Hamiltonian (\ref{harm}), where {\color {black}${\bf e}({\bf R}_{i})$  is now} the {\em local} strain at particle $i$. The probability $P(s_{\mu})$ remains Gaussian with the same variance but now the $\pm$ symmetry of $s_{\mu}$ is explicitly broken and $P(s_{\mu})$ is shifted with a mean $\langle s_{\mu} \rangle = \sum_{\bf R} {\color{black}{\bf b}^{\rm T}_{\mu}}\bar{\mathsf{C}}\mathsf{Q}^{\rm T}{\bf \Sigma} \neq 0$.
 To lowest order, therefore -- or exactly in a harmonic solid -- stress biases the distribution of defect precursors without changing their variance. Similarly, $\bf{\Sigma}$ does not affect the space-time correlation functions of $s_{\mu}$.

%
%
\section{Generating $h_X$ using laser tweezers}
\label{sec:4}
In this section, we propose an experimental realisation of the many-body term in the Hamiltonian (\ref{harmh}) using dynamic laser traps -- a technology currently available within most sophisticated experimental optics research groups. Colloidal particles are dielectric and therefore become polarised in an electric field. Fairly intense light from a laser may be used to trap these particles, which experience a force proportional to the gradient of the light intensity $I({\bf r})$ and therefore prefer to accumulate in regions of large $I({\bf r})$. This effect is extremely useful in manipulating colloidal beads in the lab to investigate myriads of  phenomena from biology to material science. There are many reviews and books on the subject, such as Ref.~\cite{HOT}. More specifically, optical traps have been used to manipulate colloidal solids, introduce defects and watch their dynamics using video microscopy~\cite{tweeze}.

The  term proportional to $h_X$ in (\ref{harmh}) depends only on the reference lattice set $\{{\bf R}\}$ and the instantaneous particle positions and can be generated for every particle $i$ once a particular configuration is known. 
For example, one can explicitly write,
\[
\chi_i = \sum_{jk} ({\bf u}_j-{\bf u}_i)^{\rm T}{\bf P}_{j-i,k-i}({\bf u}_k-{\bf u}_i)
\]
where we gather the cartesian components of $\mathsf P$ for a given pair of particles into a matrix {\color{black}${\bf P}_{\{.,.\}}$}, and assume that this matrix is zero when $j$ or $k$ are outside the neighbourhood $\Omega$ around $i$. Then
\[
NX = \sum_{ijk} ({\bf u}_j-{\bf u}_i)^{\rm T}{\bf P}_{j-i,k-i}({\bf u}_k-{\bf u}_i)
\]
The force on particle $i$ is ${\bf F}_i = - (\partial/\partial {\bf r}_i) (-{\color {black}h_X} N X)$. Direct differentiation of the expression for $X$ then gives
\begin{equation}
{\bf F}_i = 2{\color {black}h_X} \sum_{jk}
\left[
{\bf P}_{j-i,k-i}({\bf u}_i-{\bf u}_k)
+{\bf P}_{i-j,k-j}({\bf u}_k-{\bf u}_j)
\right]
\label{lforce}
\end{equation}
The first contribution comes from $\chi_i$, the second from $\chi_j$ with $j\neq i$. 
The fact that the above forces can be worked out from the positions of the particles and {\color {black}their} nearby neighbours (nearest and next-nearest neighbours, if the coarse-graining volume $\Omega$ contains exactly the nearest neighbours) suggests the following algorithm for generating a uniform non-affine field $h_X
$ for a set of $N$ colloidal particles: 
\begin{enumerate}
\item At any instant obtain the coordinates of the $N$ particles through video microscopy. 
\item Randomly choose a subset of $M$ of these particles that will have {\color {black}$h_X$}-forces applied to them.
\item For each of the $M$ particles obtain the values of the necessary forces from the coordinates of their neighbors.
\item Apply the forces by constructing a set of $M$ laser traps. The traps will need to be placed slightly away from the respective present particle positions so that the particles experience exactly the forces calculated from (\ref{lforce}). The exact displacements of the traps will depend on $I({\bf r})$ and therefore vary with the specific apparatus and implementation. 
\item In the next instant repeat steps $1-4$ above, choosing another random subset of $M$ particles to track. 
\end{enumerate}
If these steps are repeated on a time scale much faster than the typical diffusion time of colloids, then one should be able to simulate a uniform field {\color {black}$h_X$} applied across all the $N$ particles. It is possible to update dynamical traps at $200-600$ Hz, and set up at least $M = 300$ traps simultaneously for micron sized colloidal particles using {\color{black} spatial light modulator (SLM) technology}~\cite{HOT}. This should be enough to generate a uniform $h_X$ as long as the ratio of the dynamical timescale to the update timescale {\color{black}is} larger than $N/M$. Alternatively, one may also look at the effect of a local {\color {black}$h_X$} which couples to the $\chi$ of a single particle {\color{black}and can create local} defect precursors. Statistics of such local and non-uniform, dynamic, light fields may also be computed, if desired, from the formalism outlined in this work. 

\section{Discussion and conclusions}
\label{sec:5} 

In this paper we have calculated the space-time correlation functions for thermally generated non-affine fluctuations and  elastic strains in a harmonic ideal crystal. The non-affine and elastic strain fields were obtained by projecting atomic displacements into orthogonal affine and non-affine sub-spaces defined by coarse-graining over a fixed volume $\Omega$. Our results show that these correlation functions decay to zero with time and over distance although the relaxation to the late time value is oscillatory rather than monotonic. The time correlation functions for non-affine fluctuations and strains have not been described so far in the literature though we feel that they may be obtained easily for colloidal solids using video microscopy. {\color{black}This should allow verification of our results} against experimental data~\cite{zahn,kers1}. Note that the harmonic approximation that we have used throughout has been demonstrated to describe colloidal solids rather well~\cite{harm-colloid}.

{\color{black}In addition we have identified particular non-affine fluctuations in the 2d triangular lattice which, we demonstrate, are precursors to the production of dislocation- anti-dislocation pairs and arise naturally from a systematic coarse-graining procedure.} We emphasise that the defect precursors $s_{\mu}$ are {\em not} themselves defects since the equilibrium average $\langle s_\mu \rangle = 0$. 

In order to form dislocation pairs, these localised fluctuations need to {\em condense} by escaping over a, possibly stress dependent, barrier {\color{black} $\Delta f$}, a process not describable within harmonic theory\cite{chandra}. Indeed, if the bond c-d in Fig.~\ref{eigen} were to form, a Burgers circuit around particle $0$ would yield a non-zero Burgers 
vector. {\color{black} One can argue, as below, that the non-affine field $h_X$ will actually greatly enhance the formation of such dislocation dipoles in a real solid. The rate of barrier crossing is proportional to $\exp(-\beta \Delta f)$ with a prefactor, the so called ``attempt frequency'' which is a product of the characteristic frequencies of oscillation of the system in its parent state and at the saddle point~\cite{chandra}. Consider the neighbourhood $\Omega$ of a single particle. In a solid with anharmonic forces between particles, the free energy for producing a precursor fluctuation of amplitude $s_\mu$ has the form $f(s_\mu) = A s_\mu^2 - B s_\mu^4 + C s_\mu^6$, where $A, B$ and $C$ are, possibly temperature (and stress) dependent, phenomenological parameters. Note that, in the harmonic limit $B=C=0$ and $A \propto  \langle s_\mu^2 \rangle^{-1}$. This form for the free energy ensures that the $\pm s_\mu$ symmetry is preserved and a non-zero barrier for the nucleation of a dislocation dipole ($\langle s_\mu \rangle \ne 0$), given by the saddle point value of $f(s_\mu)$, exists.  When $h_X$ is turned on, this has the effect of increasing $\langle s_\mu^2 \rangle$ (see (\ref{LR-s})). This has two consequences: it decreases both the attempt frequency and $\Delta f$ with the latter effect far outweighing the former and effectively causing an overall increase in the rate of production of dislocation dipoles. For negative $h_X$, on the other hand $\Delta f$ is increased and dislocation nucleation is suppressed.} 

The dynamics considered in our formulation is entirely composed of lattice vibrations. In a crystalline solid one needs to consider, in addition, the slow vacancy diffusion mode ~\cite{martin}. Since the vacancy concentration in crystalline solids at temperatures far from melting is vanishingly small, this contribution is mostly negligible at low temperatures. However, close to the melting transition, the diffusion of vacancies does contribute significantly. Within a harmonic theory, there is, of course, no description of vacancy diffusion. On the other hand, vacancy diffusion over large distances occurs by small movements of atoms across distances of the order of the lattice spacing --- and so of the sort involved in the nucleation of a dislocation pair. Hence precursor fluctuations for vacancy diffusion may be similarly described as a non-affine distortion of a volume $\Omega$ containing a single vacancy. A calculation of vacancy migration precursors using a procedure similar to the one described in this work is in progress.  

Our calculations may also be generalised to amorphous solids. In such solids, the lack of a clearly defined reference configuration makes the identity of the relevant non-affine fluctuations debatable. The dominant deformation mechanisms in amorphous solids are local atomic rearrangements that resemble, somewhat, our defect precursor modes. However, it is not clear whether, in a particular realisation of the amorphous structure,  such precursors are ``frozen in''~\cite{manning1,manning2} even at zero stress or are produced during the deformation protocol~\cite{itamar2}. We believe that a generalisation of our calculation may be able to elucidate this point by looking at neighbourhoods with large non-affine susceptibility and determining the response to both stress and the non-affine field $h_X$.  

The results presented here may be verified in experiments on colloidal solids or dusty plasmas in the presence of a field $h_X$ produced using laser tweezers~\cite{tweeze, HOT}. Since functionalised colloidal assemblies have many technological applications, control over their structure may be of some use~\cite{colloids}.  In section~\ref{sec:4}, we outline an algorithm which, we believe, can be implemented in practice. Similar ideas have been reported in the literature~\cite{tweeze} where light fields have been used to create dislocations and grain boundaries by manipulating individual colloidal particles. We believe our approach allows greater control by targeting, instead, defect precursor fluctuations. First of all, one is able to both increase {\em as well as suppress} defect densities in a crystal by an external light field. Also, if $h_X$ is applied sufficiently slowly, the solid may be persuaded to remain in thermodynamic equilibrium {\em at a given temperature} throughout the process without producing unwanted stresses and deformations. Finally, the specific dynamics of such protocols (switching $h_X$ off or on at some rate) can be computed within the formalism discussed here. {\color{black} It is also, in principle, possible to  excite a local non-affine displacement or even a specific non-affine mode, say $s_1$, at a specific point using our ideas. For the latter case, however, one needs to know beforehand the eigenvectors of the local ${\mathsf P}{\mathsf C}{\mathsf P}$, which involves a knowledge of the interactions embodied in the ${\mathsf C}$ matrix. This introduces uncertainties that are not encountered while imposing $h_X$.}   
For dusty plasmas~\cite{colloids}, the equations we have used for the space-time correlation functions are immediately applicable. For colloidal particles dispersed in a liquid, of course, one needs to account for damping and Brownian noise terms in the dynamical equation (\ref{MATC}) to compare time-dependent quantities with experiments. Equilibrium predictions, though, would continue to be valid. Also anharmonic interactions, always present in real colloids, would lead to metastable defects {\color{black}at} positive $h_X$. Our calculations are then directly valid for small values of the field before such nucleation events actually take place. We believe that in this case, our results will be of much value for checking and validating {\color{black}the relevant} experiments. 

\begin{acknowledgments}
SG thanks CSIR India for a Senior Research Fellowship. SG and SS are grateful for funding from the FP7-PEOPLE-2013-IRSES grant no: 612707, DIONICOS. Discussions with A. Mitra, S. Karmakar and J. Horbach are gratefully acknowledged. 
\end{acknowledgments}

\end{document}